\documentclass[useAMS,usenatbib]{mnras}
\usepackage{natbib}
\usepackage{graphicx, color, url, bm}
\newcommand {\apgt} {\ {\raise-.5ex\hbox{$\buildrel>\over\sim$}}\ }
\newcommand {\aplt} {\ {\raise-.5ex\hbox{$\buildrel<\over\sim$}}\ } 

\title[Kinematics of Subclusters in Star Cluster Complexes]{Kinematics of Subclusters in Star Cluster Complexes:
Imprint of their Parental Molecular Clouds}

\author[M. S. Fujii]
{M. S. Fujii$^{1}$
\thanks{E-mail: fujii@astron.s.u-tokyo.ac.jp} 
\\
$^{1}$Department of Astronomy, Graduate School of Science, The University of Tokyo, 7-3-1 Hongo, Bunkyo-ku, Tokyo, 113-0033, Japan\\}
\begin{document}

\date{Accepted . Received ; in original form 1988 October 11}

\pagerange{\pageref{firstpage}--\pageref{lastpage}} \pubyear{2018}

\maketitle

\label{firstpage}

\begin{abstract}
Star cluster complexes such as the Carina Nebula can have formed in turbulent giant molecular clouds. We perform a series of $N$-body simulations starting from subclustering initial conditions based on hydrodynamic simulations of turbulent molecular clouds. These simulations finally result in the formation of star cluster complexes consisting of several subclusters (clumps). We obtain the inter-clump velocity distribution, the size of the region, and the mass of the most massive cluster in our simulated complex and compare the results with observed ones (the Carina Nebula and NGC 2264). 
The one-dimensional inter-clump velocity dispersion obtained from our simulations is $2.9\pm0.3$ and $1.4\pm0.4$\,km\,s$^{-1}$ for the Carina- and NGC 2264-like models, respectively, which are consistent with those obtained from Gaia Data Release 2: 2.35 and 0.99\,km\,s$^{-1}$ for the Carina Nebula and NGC 2264, respectively. 
We estimate that the masses of the parental molecular clouds for the Carina Nebula and the NGC 2264 are $4\times 10^5$ and $4\times 10^4M_{\odot}$, respectively.
\end{abstract}


\section{Introduction}

Star clusters are often born in a hierarchical structure which consists of 
several subclusters (hereafter clumps). One of the biggest systems is the Carina Nebula, 
which include several star clusters and smaller clumps 
\citep{2011ApJS..194....9F,2014ApJ...787..107K}.
Such ``star cluster complexes'' are considered to have formed via
the gravitational collapse of giant molecular clouds with turbulence
\citep[][and references therein]{2007ARA&A..45..565M}.
The formation of stars and star clusters in turbulent molecular clouds 
have been tested using numerical simulations
\citep{2008MNRAS.389.1556B,
2012MNRAS.419.3115B,2012ApJ...754...71K,2012ApJ...761..156F,2015MNRAS.449..726F,2015PASJ...67...59F,2016ApJ...817....4F}.
In these studies, hierarchical structure formation has been confirmed.

Star clusters are initially embedded 
in their parental molecular clouds \citep{2003ARA&A..41...57L},
but once massive stars formed, the gas is expelled by gas expulsion 
such as ionization, stellar winds, and supernovae explosions.
As a result, the embedded clusters are expected to expand. This
expansion has also been studied using numerical simulations
\citep{2009A&A...498L..37P,2012MNRAS.420.1503P} and also by
observations \citep{2018PASP..130g2001G,2018arXiv180702115K}.

Not only star clusters, star cluster complexes have also been suggested to 
expand by numerical simulations. 
\citet{2015MNRAS.449..726F} performed a series of simulations of star cluster 
complexes forming in turbulent molecular clouds. They suggested that 
star cluster complexes also expand although some subclusters (hereafter, clumps) merge and evolve to more massive clusters within a few Myr
\citep[see also ][]{2015PASJ...67...59F,2016ApJ...817....4F}. However,
the relative velocity among clumps in star cluster complexes was not 
studied in their work.

Observational studies on the kinematics of star cluster complexes require an accurate velocity measurement. 
Thanks to Gaia Data Release 2 \citep{2018arXiv180409365G}, 
the proper motions of individual stars in young star clusters and associations are 
now available. This data allows us to study the kinematics of star cluster complexes. 
\citet{2018arXiv180702115K} measured inter clump velocities for some star cluster complexes such as the Carina Nebula. They reported an expansion of star cluster complexes as well as that of young star clusters.

In this paper, we measure the velocity structure among clumps in star cluster complexes
using the results of our numerical simulations 
\citep{2015PASJ...67...59F,2016ApJ...817....4F} and additional new simulations. 
We connect the current spacial and velocity distributions of clumps in star cluster complexes to their parental molecular clouds and estimate the properties of the parental molecular clouds.

\section{Methods}
\subsection{Numerical Simulations}
We use the results of \citet{2016ApJ...817....4F} and also perform additional simulations for some models. 
Here, we briefly summarize the methods used in this study 
\citep[see also][]{2015PASJ...67...59F,2016ApJ...817....4F}.

First, we perform hydrodynamic simulations for molecular clouds using 
a smoothed-particles hydrodynamics (SPH) code, Fi 
\citep{1989ApJS...70..419H,1997A&A...325..972G,2004A&A...422...55P, 2005PhDT........17P} with 
the Astronomical Multipurpose Software Environment
(AMUSE) \citep{2009NewA...14..369P,2013CoPhC.183..456P,2013A&A...557A..84P}\footnote{see
  \url{http://amusecode.org/}.}.
We set-up the initial conditions of the molecular clouds using AMUSE following \citet{2003MNRAS.343..413B}.
We adopt isothermal (30K) homogeneous spheres and 
give a divergence-free random Gaussian
velocity field $\delta \bm{v}$ with a power spectrum 
$|\delta v|^2\propto k^{-4}$ \citep{2001ApJ...546..980O,2003MNRAS.343..413B}. 
The spectral index of $-4$ appears in the case of compressive turbulence (Burgers turbulence), and recent observations of molecular clouds \citep{2004ApJ...615L..45H}, and numerical simulations 
\citep{2010A&A...512A..81F, 2011ApJ...740..120R,2013MNRAS.436.1245F} 
also suggested values similar to $-4$. 
We adopt the mass and density of the molecular clouds as parameters.

In Table \ref{tb:IC}, the initial conditions of molecular clouds are summarized. The model names represent the initial mass and density; for example, m400k and d100 indicate a mass of $4\times10^5M_{\odot}$ and a density of
$100 M_{\odot}$\,pc$^{-3} \sim 1700$\,cm$^{-3}$ assuming that the mean weight per particle is $2.33m_{\rm H}$. 
We adopt 10 and $100 M_{\odot}$\,pc$^{-3}$ (170 and 1700\,cm$^{-3}$) for the density and $4\times 10^4$, $10^5$, $4\times 10^5$, and $10^6M_{\odot}$ for the mass.
While the density of our initial conditions is 170--1700\,cm$^{-3}$, observed density of molecular clouds is 100--500\,cm$^{-3}$ \citep{2015ARA&A..53..583H}.
For models m1M-d100, m400k-d100, and m400k-d10, we use the results obtained in \citet{2015MNRAS.449..726F}. We also perform simulations for additional models m100k-d100, m40k-d100, and m100k-d10.
We set the kinetic energy ($E_{\rm k}$) to be equal to the absolute value of the potential energy ($|E_{\rm p}|$). With this setting, the system is initially super-virial. For comparison, we test a model the same as model m100k-d100, but $E_{\rm k}/|E_{\rm p}|=0.5$ (model m100k-d100-vir). 

After 0.9 free-fall time of the initial condition, we stop the SPH simulations (0.75 and 2.4\,Myr for d100 and d10 models, respectively) and convert a part of the gas particles to stellar particles using the following procedure. We assume a local star formation efficiency (SFE), which depends on the local gas density $\rho$, given by 
\begin{eqnarray}
  \epsilon_{\rm loc} = \alpha_{\rm sfe} 
                       \sqrt{\frac{\rho}{100\, (M_{\odot}{\rm pc}^{-3})}},
\label{eq:eff}
\end{eqnarray}
where $\alpha_{\rm sfe}$ is a coefficient which controls the star formation
efficiency and a free parameter in our simulations. We adopt $\alpha_{\rm sfe}=0.02$.
This SFE is motivated by the 
result that the star formation rate scales with free-fall time
\citep{2012ApJ...745...69K,2013MNRAS.436.3167F}.
Using this equation, we calculate the local SFE for each gas particle. Following it, we choose gas particles which should be converted to stellar particles.
For the selected gas particles, we randomly assign stellar masses following a Salpeter mass function \citep{1955ApJ...121..161S} with a lower and upper cut-off mass of 0.3 and 100$M_{\odot}$ and converted the gas particle to a stellar particle. Although the mass does not locally conserved, the total mass globally conserves because the mean stellar mass is equal to the gas-particle mass in the SPH simulations. We assume that the velocity of the stellar particles are the same as that of their parent gas particles. With this assumption, the stellar system can take over the velocity field of their parental molecular cloud. 

Resulting global SFE measured for the entire system was several \%, but the local SFE for dense regions reaches $\sim 30$\,\%. In dense regions, the local SFE exceeds 0.5 and reaches 1.0 in the densest regions. We allow such a high local SFE following the results of previous studies, in which star formation was followed using sink particles and in dense stellar clumps stars were dominant \citep{2010MNRAS.404..721M,2012MNRAS.419..841K}.
The SFE of some models are summarized in Table 2 of \citet{2016ApJ...817....4F}.

We remove all residual gas particles and start $N$-body simulations using the stellar distribution obtained from the SPH simulations. At this moment, the virial ratio (kinetic energy over potential energy) of the entire system is more than 0.5 \citep[see Table 2 of ][]{2016ApJ...817....4F}. The entire system, therefore, expands with time. In clumps, however, stars are dominant, and as a consequence they are initially bound. Such clumps survive, and some of them evolve to more massive clusters via mergers \citep[see][for the details]{2015PASJ...67...59F}.

The $N$-body simulations are performed using a sixth-order Hermite scheme
\citep{2008NewA...13..498N} without gravitational softening.
We perform up to 10 runs changing the random seeds for the turbulence.
Runs with different random seeds result in different shapes of collapsing molecular clouds.
The number of runs and the averaged total stellar mass of each model is summarized in Table \ref{tb:clumps}.

\begin{table*}
\begin{center}
\caption{Initial Conditions of Molecular Clouds\label{tb:IC}}
\begin{tabular}{lcccc}\hline \hline
Model    & $N_{\rm Run}$ & $M_{\rm MC} (10^3M_{\odot})$  & $R_{\rm MC}$ (pc) & $\sigma_{\rm MC}$ (km\,s$^{-1}$) \\ 
\hline
m1M-d100 & 1 & 1000 & 13.3 & 19.6  \\
m400k-d100 & 3 &  400 & 10.0 & 14.4  \\
m100k-d100(-vir) & 5 & 100 & 6.2 & 9.12  \\
m40k-d100 & 10 & 40 & 4.6 & 6.70  \\
m400k-d10 & 3 & 400 & 21.0 & 9.92  \\
m100k-d10  & 6 & 100 & 13.3 & 6.23  \\
\hline
\end{tabular}
\medskip

The first column indicates the name of the model. The second column gives the number of runs. Column 3--5 are give the mass, radius, and velocity dispersion ($\sigma_{\rm MC}^2=2E_{\rm k}/M_{\rm MC}$ and $E_{\rm k}$ is the kinetic energy) of the molecular cloud. For all models we set $E_{\rm k}/|E_{\rm p}|=1.0$, but for m100k-d100-vir $E_{\rm k}/|E_{\rm p}|=0.5$. 
\end{center}
\end{table*}

\subsection{Clump finding}
At 0.5\,Myr and 2\,Myr from the beginning of the $N$-body simulation, we identify clumps in the snapshots and measure their mass and velocity. Hereafter, we set the beginning of the $N$-body simulation to be 0\,Myr.

We use the HOP method \citep{1998ApJ...498..137E} in AMUSE for the clump finding. HOP is a clump finding algorithm using peak densities. In HOP, however, the connection to the nearest densest particle is set for each particle. This is for separating multiple clumps which exist in a region denser than a threshold density. One of basic parameters of HOP is the outer cut-off density (a parameter for the minimum density threshold of clumps), $\delta_{\rm HOP}$.
We set $\delta_{\rm out} = 4.5M_{\rm s}/(4\pi r_{\rm h}^{3})$, which is three times the half-mass density of the entire system.  
We first calculate the local density of each star using $N_{\rm dens}$ nearest stars. We set $N_{\rm dense}=32$. Using the local densities, the stars are connected to the densest particles within $N_{\rm hop}$ nearest stars as a potential member of the clump. In some dense region may include multiple clumps. In such a case, the clump members can be separated to multiple clumps using the saddle density ($\delta _{\rm saddle}$). HOP finds multiple clumps using the connection and separate them using the saddle density threshold, $\delta_{\rm saddle}$. We adopt $\delta _{\rm saddle}=8\delta_{\rm out}$. We also set the peak density, $\delta_{\rm peak}=10\delta_{\rm out}$. The peak density of each clump must be higher than $\delta_{\rm peak}$. We tested some combinations of these parameters and confirmed that the inter-clump velocity dispersion does not strongly depend on the choice of the parameters. We summarize the results with different parameter sets in Appendix A.

If the mean density of the detected clump is less than $100\delta_{\rm out}$,
we repeat the same procedure for the clump, because the clump may consist of some sub-clumps. 
We set the minimum number of stars for a clump to be 50, but for models m40k-d10 and m100k-d10, we reduce it to 32 because their total mass is smaller than the other models, and as a result, the number of clumps are also small.
The detected clumps have a mass-radius relation similar to that of observed clusters. We confirmed it in our previous papers \citep{2015PASJ...67...59F,2016ApJ...817....4F}.

\begin{figure}
\begin{center}
\includegraphics[width=\columnwidth]{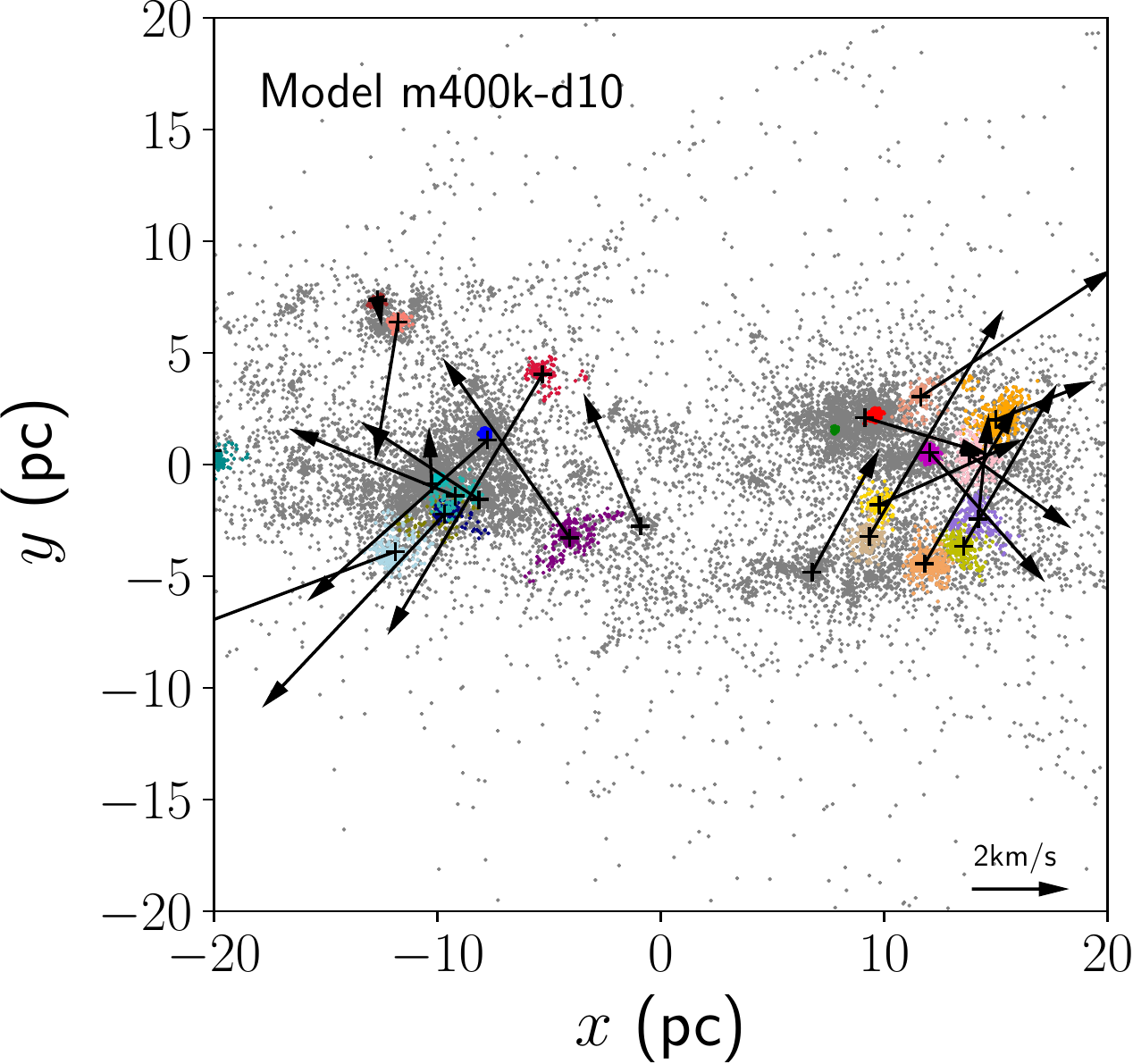}\\
\includegraphics[width=\columnwidth]{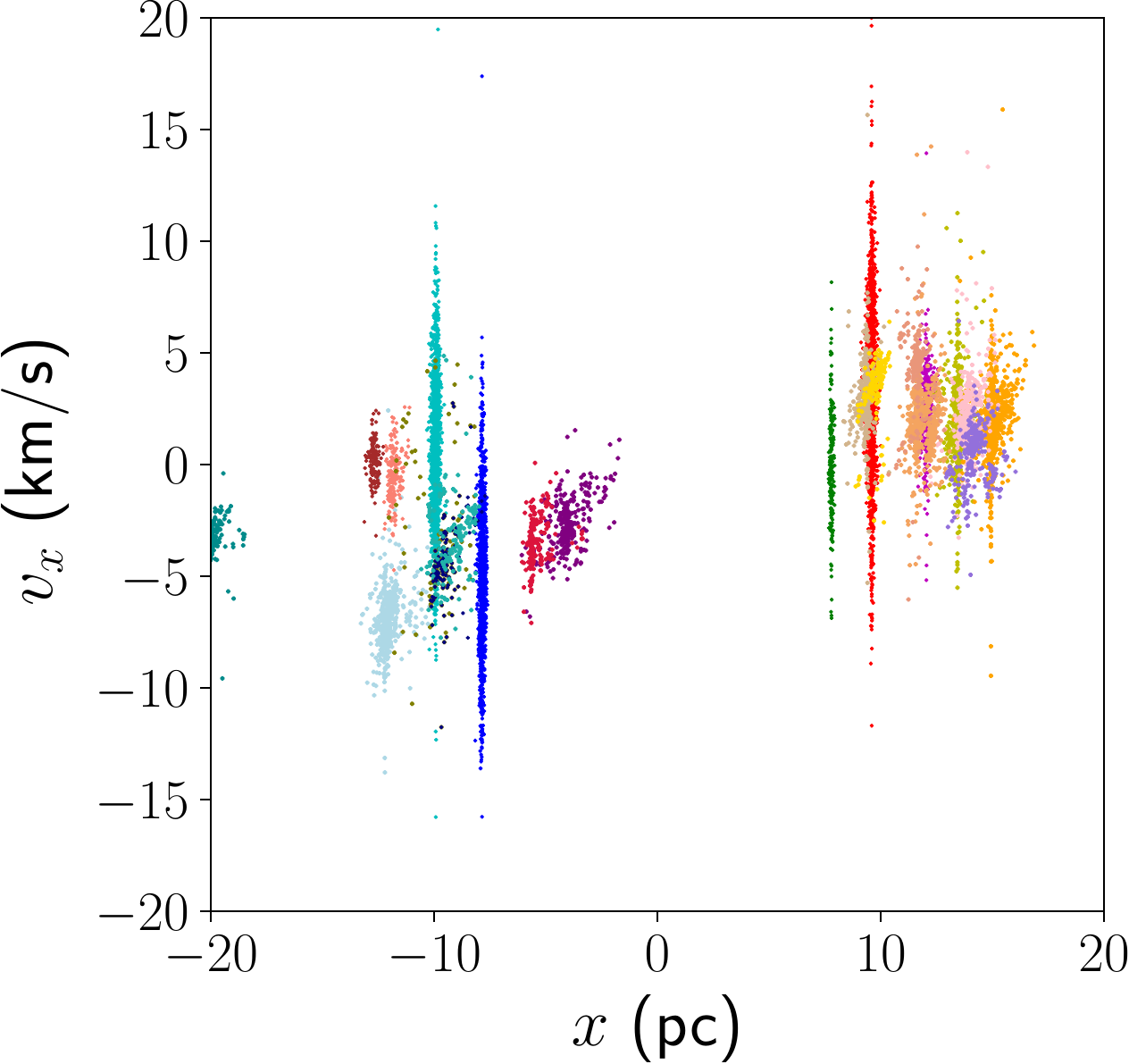}
\end{center}
\caption{Spatial and velocity distribution of stars at 0.5\,Myr for model m400k-d10, which has a mass and size distribution similar to Carina. 
Each color indicates each detected clump. Gray dots indicate the other stars.  
Arrows in the top panel indicate the velocity vector of the clumps.
\label{fig:snapshots1}}
\end{figure}

\begin{figure}
\begin{center}
\includegraphics[width=\columnwidth]{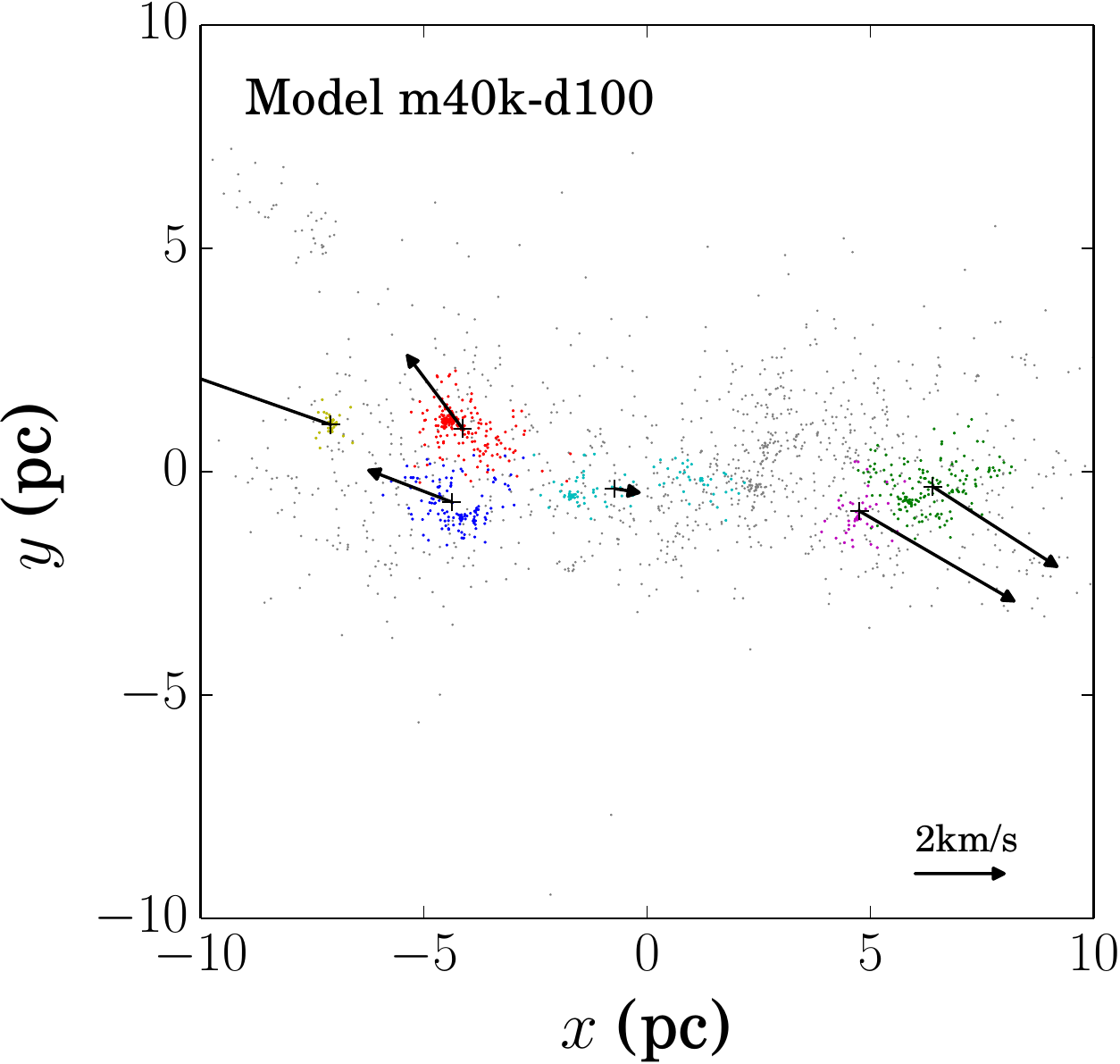}\\
\includegraphics[width=\columnwidth]{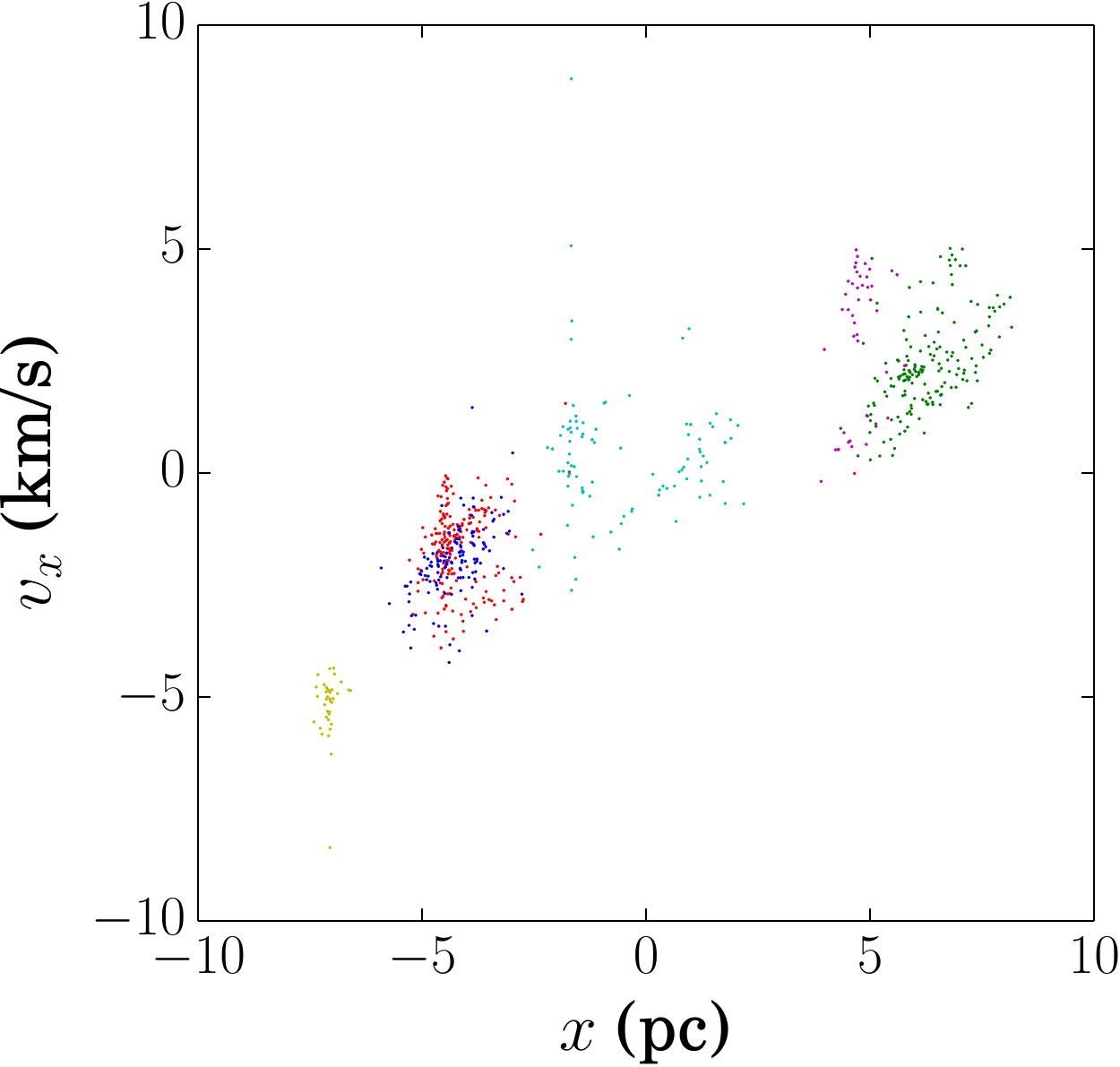}
\end{center}
\caption{Same as Fig.\ref{fig:snapshots1}, but for model
 m40k-d100 at 2\,Myr. This model has a mass and size distribution
similar to NGC 2264.
\label{fig:snapshots2}}
\end{figure}

\section{Results}

We measure the center-of-mass velocity of the detected clumps with respect to the center-of-mass velocity of the all detected clumps. 
In the top panel of Fig. \ref{fig:snapshots1}, we present the spacial distribution of stars and detected clumps for one of model m400k-d10 at 0.5\,Myr. The initial mass and density of this model are $4\times 10^5M_{\odot}$ and $10M_{\odot}$pc$^{-3}$, respectively.
We also show the velocity vector of each clump in the figure.

The averaged size (three-dimensional root-mean-square radius from the center-of-mass position) and one-dimensional velocity dispersion among clumps of this model are $r_{\rm rms}=14\pm 1$ (pc) and $\sigma_{\rm 1D}=2.9\pm0.3$\,(km\,s$^{-1}$), respectively. These values are similar to those of the Carina Nebula; the two-dimensional root-mean-square radius and the one-dimensional velocity dispersion are 9.15 pc and $2.35$\,km\,s$^{-1}$, respectively \citep{2018arXiv180702115K}. Considering the root-mean-square radius in our simulations is calculated in three dimension, the observed radius is scaled to be 11.2\,pc. The entire system of this model is distributed in $\pm 20$\,pc (see the top panel of Fig. \ref{fig:snapshots1}). The Carina Nebula is also distributed on similar scale \citep[see Fig. 13 in][]{2018arXiv180702115K}.

The relation between the mass of the parental molecular clouds ($M_{\rm MC}$) and the most massive star cluster ($M_{\rm cl, max}$) was discussed in our previous study \citep{2015MNRAS.449..726F}, and found that it follows:
\begin{eqnarray}
  \frac{M_{\rm cl, max}}{1M_{\odot}}=0.2\left( \frac{M_{\rm MC}}{1M_{\odot}}\right)^{0.76}.
\end{eqnarray}
A similar relation is also found using radiation-hydrodynamic simulations with sink particles \citep{2018NatAs...2..725H}.
Our results are roughly consistent with this relation.
The mass of the most massive cluster (clump) would be an important parameter to discuss the parental molecular clouds.
The mass of the most massive clump of model m400k-d10 is $3.3 \pm 1.9 \times 10^3M_{\odot}$. The most massive cluster in the Carina Nebula is Trumpler 14, which has a mass of $4.3^{+3.3}_{-1.5}\times 10^3M_{\odot}$ \citep{2010A&A...515A..26S}.
The total stellar mass and gas $+$ dust mass of the Carina Nebula are estimated to be $2.8\times10^4M_{\odot}$ \citep{2011A&A...530A..34P} and 
$2\times 10^5M_{\odot}$, respectively, which are similar to those of model m400k-d10; $2.5\pm 0.8\times 10^{4}M_{\odot}$ and $4\times 10^5M_{\odot}$, respectively (see Tables \ref{tb:clumps} and \ref{tb:obs}). 

In the bottom panel of Fig. \ref{fig:snapshots1}, we present the position vs. velocity plot of individual stars in the detected clumps for this model. The clumps distribute within $v_{x}\aplt|10|$\,km\,s$^{-1}$, and this is consistent with that of the Carina Nebula \citep[see Fig.12 in][]{2018arXiv180702115K}. Here, we see the entire system is expanding.

At 0.5 and 2\,Myr for each model, we calculate
the average and standard deviation of the number ($N_{\rm cl}$), one-dimensional
velocity dispersion ($\sigma_{\rm 1D}$), root-mean-square radius ($r_{\rm rms}$), and maximum mass ($M_{\rm cl, max}$) of the detected clumps among the same models with different random seeds, and these results are summarized in Table \ref{tb:clumps}. We, for comparison, summarize these values for Carina Nebula and NGC 2264 in Table \ref{tb:obs}. 
We here note that the inter-clump velocity dispersion measured in our simulations does not strongly depend on clump finding algorithms, because it is comparable to the velocity dispersion of all individual stars in the same region (see Appendix A).

While the velocity dispersion did not change much, the root-mean-square radius increased.
This expansion is due to the gas expulsion. After we removed all gas particles, the virial ratio of this system is larger than $0.5$ \citep{2015MNRAS.449..726F,2015PASJ...67...59F,2016ApJ...817....4F}. 
The velocity dispersion among clumps depends on the initial condition of the molecular clouds. Higher mass or density result in a larger velocity dispersion. We found no clear differences even if we change the initial virial ratio of the molecular clouds (see models m100k-d100 and m100k-d100-vir).
We discuss this point in Section 4.

In the top panel of Fig. \ref{fig:snapshots2}, we present the spacial distribution of clumps with their velocity vectors for one of model m40k-d100, which has a size similar to NGC 2264 region. In order to compare with the results of \citet{2018arXiv180702115K}, we also present the position vs. velocity plots for these models in the bottom panels of this figure.
In this case, we see a clear velocity gradient, which shows an expansion. 

Since the age of NGC 2264 is estimated to be $\sim 3$\,Myr 
\citep{2017A&A...599A..23V}, we compare the results of this model at 2\,Myr. The 1D velocity dispersion and root-mean-square radius at 2\,Myr is $1.4 \pm 0.4$\,km\,s$^{-1}$,
which is consistent with that of the NGC 2264 region (0.99\,km\,s$^{-1}$) \citep{2018arXiv180702115K}. 
The size ($r_{\rm rms}$) of this model is $3.2 \pm 1.5$\,pc at 2\,Myr, which is similar to that of NGC 2264 (2.53\,pc)
\citep[see Table \ref{tb:obs} and Figure 13 of][]{2018arXiv180702115K}.
Model m100k-d10 also has a velocity dispersion comparable with model m40k-d100, but the size of model m100k-d10 is $6.9 \pm 4.9$\,pc at 2\,Myr, which is twice as large as that of model m40k-d100.

We also compare the maximum mass of the most massive cluster (S Mon) in NGC 2264 with the model. In order to obtain the mass of S Mon, we use the fraction of the number of samples summarized in Table 4 in \citet{2018arXiv180702115K}. According to the table, the number of samples for S Mon is 67, and the number of all samples in NGC 2264 is 516. If we assume that the fraction in the number of samples is the same as the mass fraction of S Mon, we estimate the mass of S Mon is $150\,M_{\odot}$ from the total mass of NGC 2264 ($1100M_{\odot}$) \citep{2008A&A...487..557P}. On the other hand, the mass of the most massive clump for model m40k-d100 is $340\pm240 M_{\odot}$. The minimum value is comparable to the observation. We, therefore, estimate that NGC 2264 formed in a dense molecular cloud ($100M_{\odot}$\,pc$^{-3}$ i.e., 1700\,pc$^{-3}$) with a mass of a few $10^4M_{\odot}$.

\begin{table*}
\begin{center}
\caption{Results of simulations \label{tb:clumps}}
\begin{tabular}{lccccc}\hline \hline
Model    & $M_{\rm tot}(10^3M_{\odot})$  & $N_{\rm clump}$ & $\sigma_{\rm 1D}$(km\,s$^{-1}$) & $r_{\rm rms}$ (pc) & $M_{\rm cl,max}(10^3M_{\odot})$   \\ 
\hline
                     & \multicolumn{5}{c}{0.5\,Myr} \\
m1M-d100  & $110$ & 151 & 5.1 & 8.3 & $7.0$\\
m400k-d100  &  $31\pm 8 $ & $51 \pm 5$ & $4.0 \pm 0.9$ & $8.5 \pm 2.0$ & $3.7 \pm 3.2$ \\
m100k-d100  & $9.9 \pm 2.1$ & $8.8 \pm 4.1$ & $2.5 \pm 0.4$ & $2.4 \pm 1.1$ & $0.93 \pm 0.37$ \\
m100k-d100-vir  & $12 \pm 3$ & $13 \pm 4$ & $2.6 \pm 0.3$ & $2.5 \pm 0.7$ & $1.1 \pm 0.9$ \\
m40k-d100  & $2.3\pm 0.5$ & $5.6 \pm 1.4$ & $1.6 \pm 0.3$ & $2.1 \pm 1.0$ & $0.47\pm 0.27$ \\
m400k-d10  &  $25 \pm 8$ & $47 \pm 6$ & $2.9 \pm 0.3$ & $14 \pm 1$ & $3.3 \pm 1.9$ \\
m100k-d10  &  $3.1 \pm 1.5$ & $7.2 \pm 3.4$ & $1.6\pm0.2$ & $7.1\pm 3.6$ &  $0.51 \pm 0.39$  \\
\hline
               & \multicolumn{5}{c}{2\,Myr} \\
m1M-d100  & $110$ & 108 & 4.5 & 18 & $7.7$\\
m400k-d100  &  $31\pm 8$ & $37\pm 9$ &  $3.7 \pm 0.9$ & $17 \pm 5$ & $4.4 \pm 3.9$ \\
m100k-d100  & $9.9 \pm 2.1$ & $8.2 \pm 4.3$ & $2.0 \pm 0.5$ &  $7.2 \pm 1.9$ & $1.5 \pm 1.1$ \\
m100k-d100-vir  & $12 \pm 3$ & $13 \pm 5$ & $2.3 \pm 0.5$ &  $7.7 \pm 1.8$ & $2.3 \pm 2.2$ \\
m40k-d100  & $2.3\pm 0.5$ & $4.4 \pm 0.4$ & $1.4 \pm 0.4$ & $3.2 \pm 1.5$ & $0.34 \pm0.24$ \\
m400k-d10  &  $25 \pm 8$ & $44 \pm 12$ &  $2.7 \pm 0.2$ & $16 \pm 0.4$ & $2.2 \pm 1.2$ \\
m100k-d10   &  $3.1 \pm 1.5$ &  $5.5 \pm 2.9$ & $1.5 \pm 0.9$ & $6.9 \pm 4.9 $ & $0.41 \pm 0.43$  \\
\hline
\end{tabular}
\medskip

\end{center}
\end{table*}

\begin{table*}
\begin{center}
\caption{Observed star cluster complexes \label{tb:obs}}
\begin{tabular}{lccccccc}\hline \hline
Name    & Age & $M_{\rm tot}(10^3M_{\odot})$ & $N_{\rm clump} $ & $\sigma_{\rm 1D}$(km\,s$^{-1}$) & $\sqrt{3/2}\,r_{\rm rms,2D}$ (pc) & $M_{\rm cl, max}(10^3M_{\odot})$ & Ref. \\ 
\hline
Carina  & 0.5 & 28 & 16 & 2.35 & 11.2 & 4.3 & (1), (2), (3)\\
NGC 2264 & 3 & 1.1 & 8 & 0.99 & 2.53 & 0.15 & (3), (4), (5) \\
\hline
\end{tabular}
\medskip

(1) \citet{2010A&A...515A..26S}; (2) \citet{2011A&A...530A..34P}; (3) \citet{2018arXiv180702115K}; (4) \citet{2017A&A...599A..23V}; (5) \citet{2008A&A...487..557P}
\end{center}
\end{table*}

In our method, we assumed an instantaneous gas expulsion. In observed star cluster complexes, however, the gas mass is comparable to or large than stellar mass. In the Carina Nebula, for example, the estimated gas mass including dust is $2\times10^5M_{\odot}$ \citep{2011A&A...525A..92P}, which is an order of magnitude larger than that of stellar mass, $2.8\times10^{4}M_{\odot}$ \citep{2011A&A...530A..34P}. 
We, therefore, may overestimate the inter-cluster velocity dispersion in our simulations.

\section{Discussion}

\begin{figure}
\begin{center}
\includegraphics[width=\columnwidth]{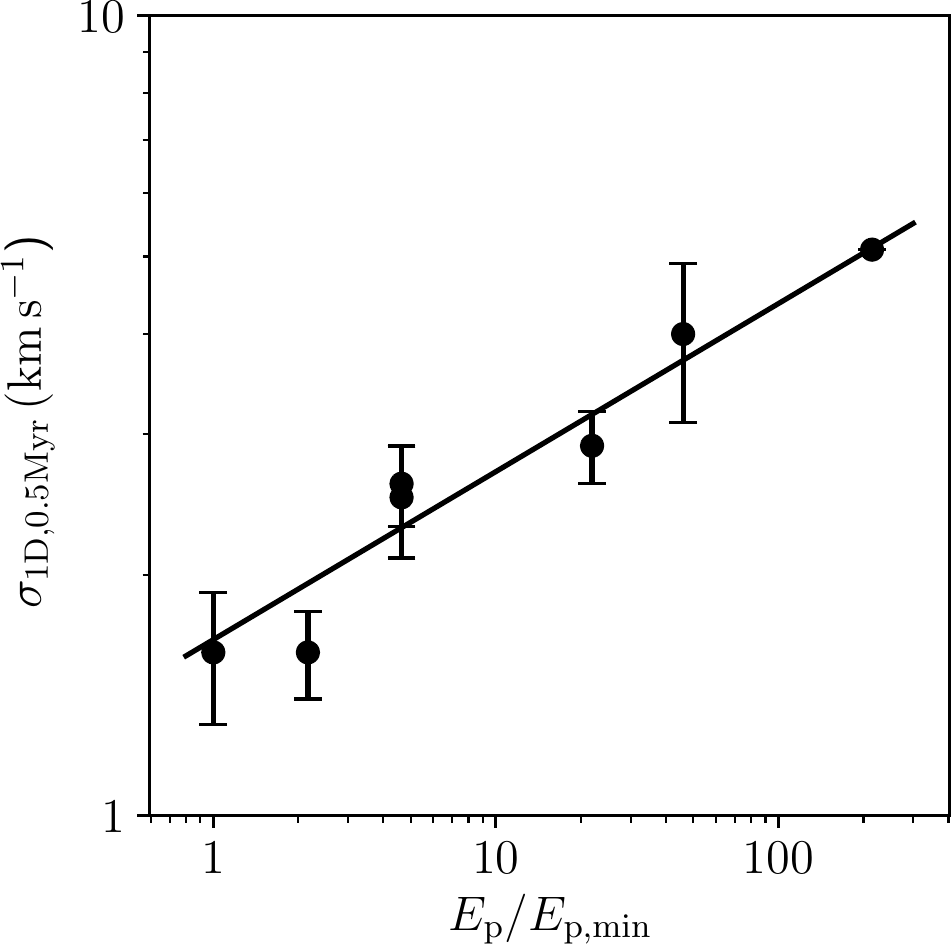}
\end{center}
\caption{Relation between the potential energy of the initial molecular clouds and the inter-clump velocity dispersion at 0.5\,Myr. The black line show the result of a least-mean-square fitting; $\sigma_{\rm 1D, 0.5Myr}=1.66(E_{\rm p}/E_{\rm p, min})^{0.21}$. Here, $E_{\rm p, min}$ is the potential energy of model m40k-d100.
\label{fig:E_vel_rel}}
\end{figure}

Which initial parameter decides the inter-cluster velocity dispersion? In our simulations, the velocity dispersion depends on the potential energy of the initial molecular cloud.
In Fig. \ref{fig:E_vel_rel}, we present the relation between the potential energy of our initial molecular clouds and the inter-cluster velocity dispersion at 0.5\,Myr. Since the velocity dispersion does not change much at 2\,Myr, we fit a power-law function to this relation using a least-mean-square method and obtain $\sigma_{\rm 1D, 0.5Myr}=1.66(E_{\rm p}/E_{\rm p, min})^{0.21}$ (km\,s$^{-1}$), where $E_{\rm p, min}$ is the minimum $E_{\rm p}$ among our models; specifically, $E_{\rm p}$ for model m400k-d10.

\begin{figure}
\begin{center}
\includegraphics[width=\columnwidth]{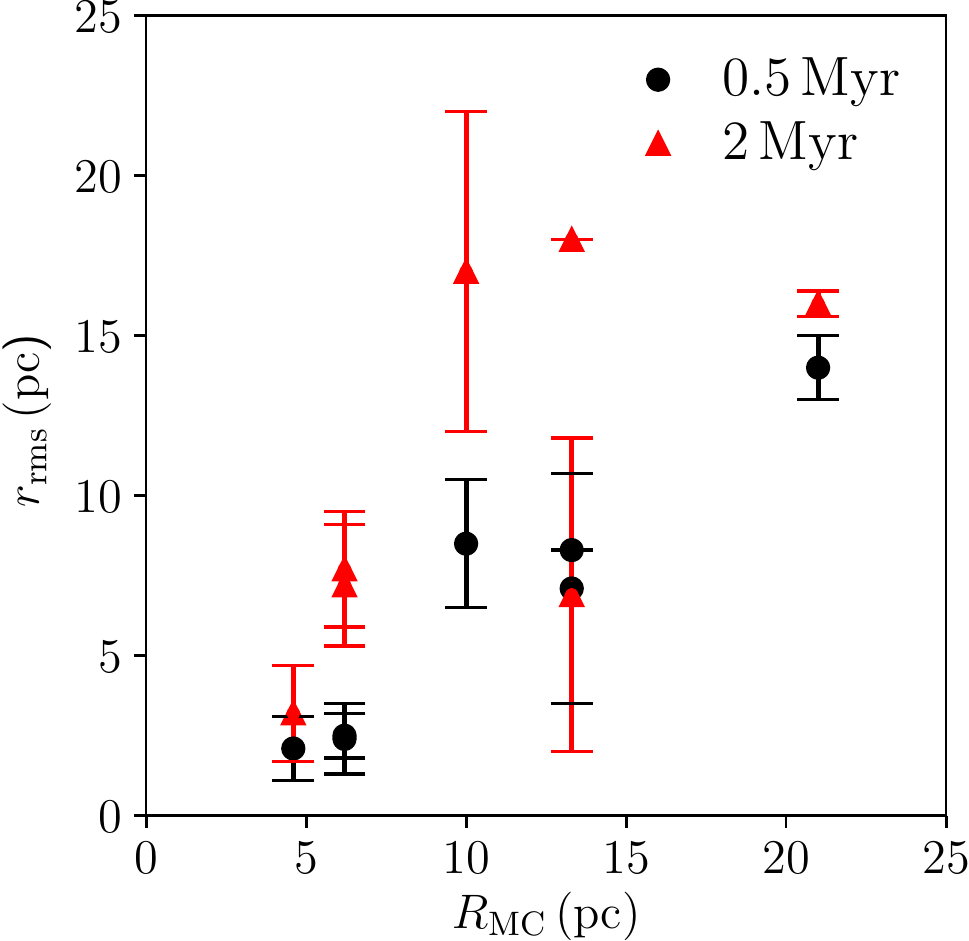}
\end{center}
\caption{Relation between the radius of the initial molecular clouds and the root-mean-square radius of the resulting star cluster complexes at 0.5\,Myr and 2\,Myr. 
\label{fig:cl_size}}
\end{figure}

In Fig. \ref{fig:cl_size}, we plot the relation between the initial size of the molecular cloud and the size of the resulting star cluster complexes at 0.5 and 2\,Myr.  At 0.5\,Myr, the sizes of the complexes are correlated with those of the initial molecular cloud, but not in later time. This is because the expansion velocity of the complexes depends on the potential energy of the initial molecular cloud. Even if the initial size of the molecular cloud is the same, the expansion velocity can be different comparing models with different densities (see models m1M-d100 and m100k-d10). 

In our study, we tested only initially spherical models. In the Orion A molecular cloud, however, 
stellar and proto-stellar clumps including the Orion Nebula Cluster is associated with a 50-pc scale filament \citep{2016AJ....151....5M,2018AJ....156...84K}.
In such a region, the initial molecular cloud might have been cylindrical \citep{2008MNRAS.389.1556B}, or a cloud-cloud collision might have triggered the star cluster formation \citep{2018ApJ...859..166F}.
In a large velocity dispersion of stars around the Integral Shaped Filament \citep{1987ApJ...312L..45B} associated with the Orion Nebula Cluster \citep{2016A&A...590A...2S}, a magnetic field may play an important role to eject stars from the filament by the ``slingshot" mechanism \citep{2017MNRAS.471.3590B,2018MNRAS.473.4890S}.

\section{Summary}

We performed a series of $N$-body simulations for the formation of star cluster complexes. Following the method in \citet{2015MNRAS.449..726F}, we first performed SPH simulations of turbulent molecular clouds and then used the last snapshots to generate initial conditions for the $N$-body simulations, in which stars are distributed in clumpy and filamental structures. 

The one-dimensional inter-clump velocity dispersion obtained from our simulations is $2.9\pm0.3$ and $1.4\pm0.4$\,km\,s$^{-1}$ for the Carina- and NGC 2264-like models, respectively, which are consistent with those obtained from Gaia Data Release 2, which are 2.35 and 0.99\,km\,s$^{-1}$ for the Carina Nebula and NGC 2264. The simulated complexes expand with time. 
We also confirmed the size and the mass of the most massive clump in these models are consistent with the observations.

Our results suggest that the parental molecular cloud of NGC 2264 has a mass of $\sim 4\times 10^4M_{\odot}$ and that Carina Nebula formed from a giant molecular cloud with a mass of $\sim 4\times10^5M_{\odot}$, but the cloud density for NGC 2264 is estimated to be higher than that of the Carina Nebula.
The inter-cluster velocity dispersion in our simulations, however, tends to be larger than that of observed star cluster complexes. This may be because we assumed an instantaneous gas expulsion, while observed star cluster complexes are still surrounded by molecular gas comparable or more massive than the total stellar mass.

\section*{Acknowledgments}
The author thanks the referee, Richard Parker, for his useful comments.
The author also thanks Jun Makino, Amelia Stutz, and Tjarda Boekholt for fruitful discussion.
Numerical computations were carried out on Cray
XC30 and XC50 CPU-cluster at the Center for Computational Astrophysics (CfCA) of the National Astronomical Observatory of Japan. The author was supported by The University of Tokyo Excellent Young Researcher Program.
This work was supported by JSPS KAKENHI Grant Number 26800108 and 19H01933.

\bibliographystyle{mnras}
\bibliography{reference}

\appendix
\section{Clump finding algorithm: HOP}
We determined subclusters using HOP algorithm \citep{1998ApJ...498..278E} in AMUSE. HOP is a clump finding algorithm based on density \citep[e.g.,][]{1991ComPh...5..164B}. In Friends-of-Friends (FoF) methods, multiple clumps connected with a bridging region can be detected as a clump \citep[see e.g.,][]{2018MNRAS.481.1679P}. In order to avoid this problem, in HOP, the connection to the nearest densest particle is set for each particle. This determine a local density peak to which each particle belongs.

Here, we briefly summarize how HOP works. HOP first calculates a local density of each particle using $N_{\rm dens}$ nearest particles, where $N_{\rm dens}$ is a parameter. The local density is determined by a spherical cubic spline kernel \citep{1985A&A...149..135M}. The recommended value in \citet{1998ApJ...498..278E} is $N_{\rm dens}=64$. 
Next, the densest particle within $N_{\rm hop} -1$ nearest particles for each particle is determined. Here, $N_{\rm hop}$ is a parameter, and the recommended value in \citet{1998ApJ...498..278E} is $N_{\rm hop}=16$. With this process, each particle is connected to a local density peak which the particle belongs to. 

For the density, there are three parameters; peak density ($\delta_{\rm peak}$), saddle density ($\delta_{\rm saddle}$), and outer density ($\delta_{\rm outer}$) thresholds. Density peaks higher than $\delta _{\rm peak}$ are detected as individual clumps. On the other hand, particles with a density lower than $\delta_{\rm outer}$ are excluded from clumps. Particles with a local density higher than $\delta_{\rm outer}$ are clump candidates. 

In regions with a density higher than $\delta_{\rm outer}$, several clumps can be included. In order to detect such clumps, $N_{\rm merge}$ is used. For a star in a density peak, if one of its $N_{\rm merge}$ nearest particles belongs to another density peak, then the averaged density of the two density peaks is calculated. If the averaged density is less than saddle density ($\delta _{\rm saddle}$), these two density peaks are detected as two clump. If not, they are treated as one clump. The recommended value for $N_{\rm merge}$ is 4.

The peak density ($\delta_{\rm peak}$) and saddle density $\delta_{\rm saddle}$ are determined depending on $\delta_{\rm outer}$. In \citet{1998ApJ...498..278E}, $\delta_{\rm peak}=3\delta_{\rm outer}$ and $\delta_{\rm saddle}=2.5\delta_{\rm outer}$ are recommended. The outer density threshold ($\delta_{\rm outer}$) should be chosen for each system, and can significantly change the result. 
In our study, we adopt three times of the half-mass density of the system (the density within a radius in which the half of the total mass is includes) as $\delta_{\rm outer}$. The half-mass density in our models was typically 10--100 $M_{\odot}$\,pc$^{-3}$ at 0.5\,Myr and 1--10 $M_{\odot}$\,pc$^{-3}$ at 2\,Myr.
For $\delta_{\rm peak}$ and $\delta_{\rm saddle}$, we adopt $\delta_{\rm peak}=8\delta_{\rm outer}$ and $\delta_{\rm saddle}=10\delta_{\rm outer}$, which values are higher than those recommended in \citet{1998ApJ...498..278E}. In \citet{1998ApJ...498..278E}, they applied this method to a cosmological $N$-body simulations and chose the parameters. In our simulations, however, the density contrast in star cluster complexes would be higher than that of dark matter halos. We, therefore, chose a higher values for $\delta_{\rm peak}$ and $\delta_{\rm saddle}$.
Even if we adopt density thresholds same as those adopted in \citet{1998ApJ...498..278E}, the results did not change significantly (see Table \ref{tb:hop}).

\begin{figure*}
\begin{center}
\includegraphics[width=\columnwidth]{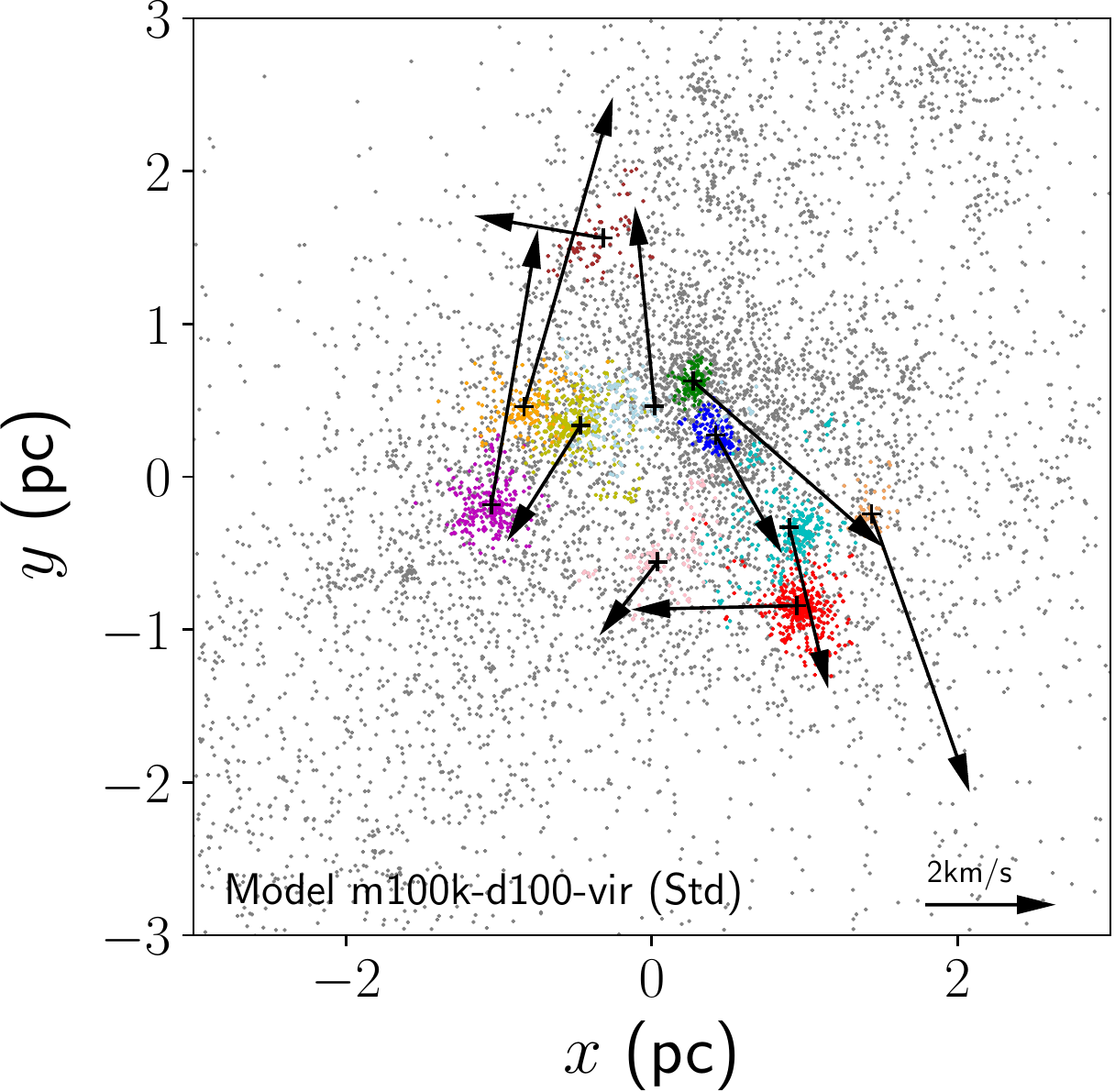}
\includegraphics[width=\columnwidth]{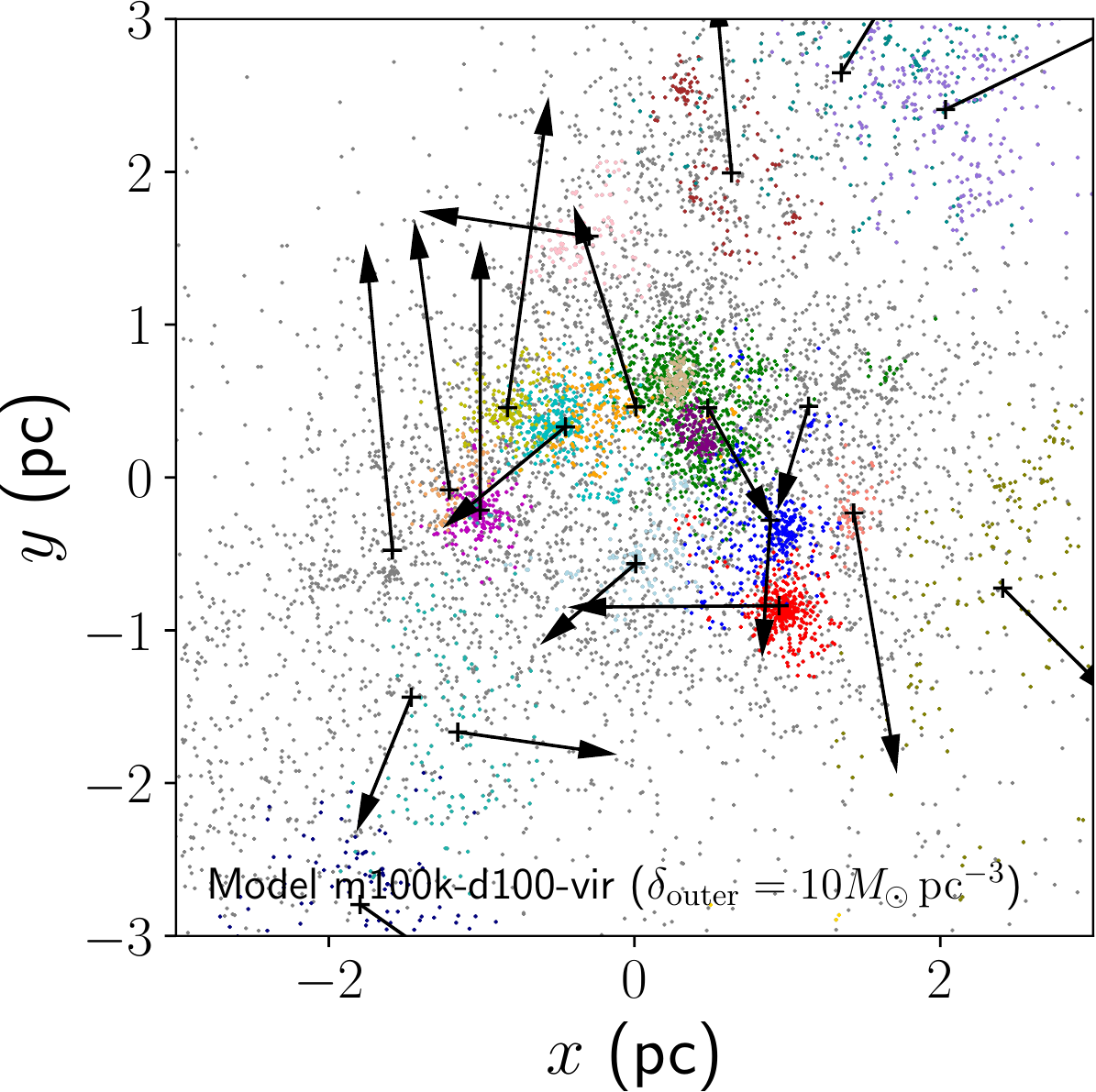}
\end{center}
\caption{Snapshots at 0.5 Myr for one of m100k-d100-vir. Each color indicate each detected clump. Gray dots indicate stars which do not belong to clumps. Arrows show the center-of-mass velocity of the detected clumps. Left: with our standard parameter set for HOP. Right: with outer density threshold ($\delta_{\rm outer}$) of $10M_{\odot}$\,pc$^{-3}$, but standard values for the other parameters.
\label{fig:appdx}}
\end{figure*}

We also tested a fixed value for $\delta_{\rm outer}$. If we adopted $\delta_{\rm}=10M_{\odot}$\,pc$^{-3}$ for both 0.5 and 2\,Myr, the number of clumps at 0.5\,Myr was twice as large as that obtained using our standard setting. However, the velocity dispersion among the detected clumps was not much different from that obtained using the standard setting. In Figure \ref{fig:appdx}, we present the positions and velocities of detected clumps with our standard setting (left) and $\delta_{\rm}=10M_{\odot}$\,pc$^{-3}$ (right). In the right panel, we see that a larger number of clumps with lower densities are identified because of the lower density threshold compared with our standard setting. We also find that some of the detected clumps in the right panel would not be identified as clumps by eye. Thus, a fixed outer density threshold for all models and all time does not work well, and we therefore adopt the mean density of the entire system as the outer density threshold rather than a fixed one. 

For relatively large clumps, however, the determined velocities are consistent with those obtained from our standard settings. We also confirmed that the inter-clump velocity dispersion does not largely change even if we change the detection criteria. This is because the inter-clump velocity dispersion is similar to the velocity dispersion of all individual stars in the same region. We calculated the velocity dispersion within the $r_{\rm rms}$, where we set it to be 3\,pc. The velocity dispersion of all stars within 3\,pc from the center of mass position of the complex is $3.4\pm 0.4$\,km\,s$^{-1}$ at $t=0.5$\,Myr. At $t=2$\,Myr, the velocity dispersion of all stars within $r_{\rm rms}(\equiv 6$\,pc) is $2.0\pm 0.4$\,km\,s$^{-1}$. Thus, inter-clump velocity dispersion is an index relatively independent of clump finding algorithms.

We also changed the values of $N_{\rm dens}=32$ and $N_{\rm hop}=32$ from the recommended values in \citet{1998ApJ...498..278E}. In our standard setting, we reduced the value of $N_{\rm dens}$ in order to detect clumps consisting of less than 50 particles. On the other hand, we increased $N_{\rm hop}$ to 32 in order to search peaks in slightly wider range of particles. However, the choice of $N_{\rm hop}$ and $N_{\rm dens}$ does not affect the results. In Table \ref{tb:hop}, we summarize the results using different values for HOP parameters.

\begin{table*}
\begin{center}
\caption{Results of m100k-d100 using different HOP parameters \label{tb:hop}}
\begin{tabular}{lcccc}\hline \hline
Changed parameters  & $N_{\rm clump}$ & $\sigma_{\rm 1D}$(km\,s$^{-1}$) & $r_{\rm rms}$ (pc) & $M_{\rm cl,max}(10^3M_{\odot})$   \\ 
\hline
     & \multicolumn{4}{c}{0.5\,Myr} \\
Standard & $13 \pm 4$ & $2.6 \pm 0.3$ & $2.5 \pm 0.7$ & $1.1 \pm 0.9$ \\
$\delta_{\rm peak}=3\delta{\rm outer}$ & $11 \pm 3.2 $ & $2.7 \pm 0.31 $ & $2.7 \pm 1 $ &
$0.78 \pm 0.38 $\\
$\delta_{\rm outer}=10 M_{\odot}$\,pc$^{-3}$ & $22 \pm 4 $ & $2.9 \pm 0.21 $ &  $3.3 \pm 0.66 $ & $1.9 \pm 1.5 $ \\
$N_{\rm hop}=8$ & $12 \pm 4.5 $ & $2.6 \pm 0.23 $ & $2.6 \pm 0.68 $ & $1.1 \pm 0.87 $ \\
$N_{\rm hop}=64$ & $13 \pm 4.5 $ & $2.6 \pm 0.24 $ & $2.6 \pm 0.8 $ & $1.2 \pm 0.88 $ \\
$N_{\rm dens}=16$ & $14 \pm 5.2 $ &  $2.7 \pm 0.18 $ & $2.6 \pm 0.8 $ & $1.2 \pm 0.89 $ \\
$N_{\rm dens}=64$ & $12 \pm 3.5 $ & $2.6 \pm 0.18 $ & $2.4 \pm 0.35 $ & $1.1 \pm 0.87 $ \\
\hline

  & \multicolumn{4}{c}{2\,Myr} \\
Standard & $13 \pm 5$ & $2.3 \pm 0.5$ &  $7.7 \pm 1.8$ & $2.3 \pm 2.2$ \\
$\delta_{\rm peak}=3\delta{\rm outer}$ & $12 \pm 4 $ & $2.2 \pm 0.38 $ & $7.5 \pm 1.6 $ & $1.6 \pm 0.97 $ \\
$\delta_{\rm outer}=10 M_{\odot}$\,pc$^{-3}$ & $10 \pm 3 $ & $2.2 \pm 0.52 $ & $7.5 \pm 2 $ &
$1.1 \pm 0.58 $\\
$N_{\rm hop}=8$ & $13 \pm 4.8 $ & $2.2 \pm 0.46 $ & $6.4 \pm 2.5 $ & $1.3 \pm 0.84 $ \\
$N_{\rm hop}=64$ & $13 \pm 5.4 $ & $2.3 \pm 0.51 $ & $7.8 \pm 2 $ & $2.4 \pm 2.2 $ \\
$N_{\rm dens}=16$ &$13 \pm 3.8 $ & $2.3 \pm 0.53 $ &  $6.5 \pm 2.9 $ &  $2.3 \pm 2.1 $ \\
$N_{\rm dens}=64$ &  $11 \pm 4.3 $ & $2.3 \pm 0.46 $ & $7.6 \pm 1.8 $ & $2.4 \pm 2.2 $ \\
\hline
\end{tabular}
\medskip

\end{center}
\end{table*}

\label{lastpage}

\end{document}